\documentclass[%
 aip,
 jmp,%
 amsmath,amssymb,
 reprint,%
]{revtex4-1}
\title{Brownian Motors and Stochastic Resonance} 
\usepackage{graphicx}
\usepackage{dcolumn}
\usepackage{bm}

\begin{document}
\preprint{AIP/123-QED}

\title{Brownian motors and stochastic resonance}

\author{Jos\'e L. Mateos}\email{mateos@fisica.unam.mx}
\affiliation{ Instituto de F\'{\i}sica, Universidad Nacional Aut\'onoma de M\'exico, Apartado Postal 20-364, 01000 M\'exico,
D.F., M\'exico.}%

\author{Fernando R. Alatriste}%
 
\affiliation{ Maestr\'{\i}a en Din\'amica No Lineal y Sistemas Complejos, Universidad Aut\'onoma de la Ciudad de M\'exico, C.P. 03100 M\'exico D.F., M\'exico}%

\date{\today}
\begin{abstract}
 We study the transport properties for a walker on a ratchet potential. 
The walker consists of two particles coupled by a bistable potential that
allow the interchange of the order of the particles while moving through
a one-dimensional asymmetric periodic ratchet potential. 
We consider the stochastic dynamics of the  walker on a ratchet 
with an external periodic forcing, in the overdamped case.
The coupling of the two particles corresponds to a single effective particle,
describing the internal degree of freedom, in a bistable potential. This 
double-well potential is subjected to both a periodic forcing and noise,
and therefore is able to provide a realization of the phenomenon of stochastic
resonance. The main result is that there is an optimal amount of noise 
where the amplitude of the periodic response of the system is maximum, 
a signal of stochastic resonance, and that precisely for this optimal noise the 
average velocity of the walker is maximal, implying a strong link between 
stochastic resonance and the ratchet effect.  
                           
\end{abstract}
\maketitle

{\bf
\noindent Nowadays is well known that noise, instead of being an annoying feature that has to be removed, 
can play a constructive role for some nonlinear systems. Two important phenomena in this
category are: Brownian motors and stochastic resonance. The former refers to the effect where
an asymmetry in a nonlinear system can rectify an unbiased fluctuating force
out of equilibrium, thereby generating a directional current; the latter refers to the effect where
a weak coherent input subthreshold signal in a nonlinear system can be detected with the 
assistance of noise. It seems that these two effects must be related, however it is not clear
under which circumstances this relation can be revealed. In this paper we present a model
that establish such a connection. The model is inspired by the dynamics of motor proteins,
like Kinesin or Myosin, and consists of a walker comprising two particles or Brownian motors,
coupled through a bistable double-well potential, that walks through an asymmetric ratchet. 
For this model we show that for an optimal amount of noise the current
generated by the ratchet effect is maximal and that this optimal noise intensity corresponds 
precisely to the maximal response characterizing stochastic resonance. Therefore, our model establishs
 a link between both phenomena.}

\section{Introduction}

In this special issue we are paying homage to our dear friend Frank Moss, 
who throughout his long and productive carrier has contributed with a series 
of seminal papers in many different fields: biological physics,  
stochastic dynamics, condensed matter, stochastic resonance, to name but a few. 
In particular, his ideas and seminal experiments on different organisms were
crucial to establish the presence of the phenomenon of stochastic resonance in 
the biological realm. His ideas and influence will endure and will be the inspiration 
for other generation of scientist. We will miss his cheerful and wit character that
manage to shape this area of research and built a strong community around his
leadership. 

This paper is devoted to Brownian motors and stochastic resonance and is dedicated to 
Frank Moss as a humble tribute. 
 
Brownian motors (or thermal ratchets) are nonlinear systems that can extract usable
work from unbiased non-equilibrium fluctuations. The canonical example is a 
particle undergoing a random walk in a periodic asymmetric (ratchet) potential, and 
being acted upon by an external time-dependent force of zero average. 
The recent burst of work is motivated, for instance, by the challenge to model
unidirectional transport of molecular motors within the biological
realm and the potential for novel applications that enables an efficient scheme to shuttle, 
separate and pump particles on the micro- and even nanometer scale 
\cite{rev2009}. In particular, it is worth mentioning the focus issue 
on ``The constructive role of noise in fluctuation driven transport and stochastic resonance'', 
in this very journal, edited by Dean Astumian and Frank Moss \cite{mosschaos}. 

In this paper, we will study the transport properties of a walker (dimer) moving through a
ratchet potential. This walker comprises two point particles coupled through a 
nonlinear force modeled by a bistable potential that opens the possibility of exchanging
the order of the two particles while walking. This model was inspired by 
the physics of molecular motors, in particular Kinesin or Myosin, which are motor proteins that
has two portions acting as ``feet'' that move through microtubules inside cells\cite{mat2005}. 
Other authors have explored different models of Brownian walkers, like the model of Brownian steepers presented in \cite{prager2005}, where each step is composed of two processes: an activation process describing the random attachment of a fuel molecule, followed by a conformational change of the stepper that leads to a forward unidirectional motion. The time-periodic modulation of the rate of the fuel concentration allows improving and regularizing the random motion of the steeper.
For a recent review of the physics of molecular motors see \cite{ajp2009}. 
Further aspects of the model discussed in \cite{mat2005} has been explored recently 
by other autors \cite{pla2006,physA2007,wio2008,epl2009,reviran}. 

We will study the dynamics, in the overdamped regime, of the above-mentioned walker, 
in a ratchet potential. Each of the two particles is characterized by a
coordinate that obeys a Langevin equation that includes
thermal noise, the force due to the ratchet potential, an external periodic force, and a
coupling force between the two particles. This coupling is given by a bistable 
(double-well) potential in terms of an internal degree of freedom given by the difference 
of the coordinates of the two particles. Therefore, this internal degree of freedom 
experience a double well potential subjected to a periodic forcing and noise, and thus 
is able to manifest the phenomenon of stochastic resonance.   

Stochastic resonance  involves the interplay of nonlinearity and
noise, in which a signal detection can be amplified and optimized by the assistance
of noise. It involves the following essential features: an energetic barrier
or threshold, a coherent (periodic) signal, and noise 
\cite{revsr1,revsr2,revsr3,revsr4,revsr5,revsr6,revsr7,revsr8,revsr9}. 
The field of stochastic resonance began in the early eighties and start 
to flourish in the nineties, specially due to the breakthrough experiments of 
Frank Moss and collaborators, that show the effect of stochastic resonance in living 
organisms \cite{moss1,moss2,moss3,moss4}.

The paradigmatic model is a particle in a bistable double well potential 
that is rocked periodically and subjected to stochastic noise. 
In the absence of noise and for a weak periodic signal, the particle is in a
subthreshold regime and cannot visit the two wells in the double well potential,
being confined to only one of them. But, with the aid of a noisy signal,
the particle is now capable of surmounting the potential barrier and start to
visit both wells. If the noise is very weak, this particle
remains confined in one of the minima and for a very strong noise the resulting
dynamics becomes fully stochastic, masking the periodic signal. However,
between these two extremes, there is an optimal amount of noise where the 
dynamics fully reveals the subthreshold periodic signal. That is, noise helps
to detect the otherwise hidden periodicity. This is, in a nutshell, the phenomenon
of stochastic resonance.  

\section{A walker with two Brownian motors and stochastic resonance}

The model considers a walker moving on an asymmetric ratchet that has 
two feet that are represented by two particles coupled nonlinearly through a
bistable potential \cite{walker031,walker032,fnl2004,mat2005,mat2010}. Additionally,
there is an external periodic driving and thermal noise. 
The equations of motion for the two particles, 
represented by $x$ and $y$, in the overdamped regime, are

\begin{equation}
m\gamma \dot x = - {\frac{dV(x)}{dx}} - {\frac{\partial V_{b}(x-y)}{\partial x}} + 
m\gamma \sqrt{2D}\xi_{1}(t) - F_D \cos(\omega_D t),
\end{equation}

\begin{equation}
m\gamma \dot y = - {\frac{dV(y)}{dy}} - {\frac{\partial V_{b}(x-y)}{\partial y}} + 
m\gamma \sqrt{2D}\xi_{2}(t) + F_D \cos(\omega_D t),
\end{equation}

\noindent where $m$ is the mass of the particle, $\gamma$ is the friction coefficient,
$V(x)$ is the asymmetric periodic ratchet potential, $V_{b}(x-y)$ is the bistable potential,
and $F_D$ and $\omega_D$ represent the amplitude and the 
frequency of the external driving force, respectively. 

These equations represent two coupled particles on a ratchet potential, which is given by

\begin{equation}
V(x) = V_{1} - V_{r} \left [\sin {\frac{2\pi (x-x_0)}{L}} + {\frac{1}{4}}\sin {\frac{4\pi (x-x_0)}{L}} \right ],
\end{equation}

\noindent where $L$ is the periodicity of the potential, 
$V_{r}$ is the amplitude, and $V_{1}$ is an arbitrary constant. The potential
is shifted by an amount $x_0$ in order that the minimum of the potential
is located at the origin \cite{ma1,ma2008}.

The bistable potential is given by

\begin{equation}
V_{b}(x-y) = V_{b} + V_{b} \left [{\frac{(x-y)^4}{l^4}} - 2{\frac{(x-y)^2}{l^2}} \right ].
\end{equation}
Here, $V_{b}$ is the amplitude of the bistable potential and represents the coupling strength
between the particles, and $2l$ is the distance between the two minima.

Finally, the parameter $D$ is the intensity of the zero-mean statistically independent
Gaussian white noises $\xi_{1}(t)$ and $\xi_{2}(t)$ acting on particles $x$ and $y$, 
respectively. Being statistically independent, they satisfy:

\begin{equation}
\langle \xi_{i}(t) \xi_{j}(s) \rangle = \delta_{ij} \delta(t-s). 
\end{equation}

Let us define the following dimensionless units: 
$x^{\prime }=x/L$, $x_{0}^{\prime }=x_{0}/L$, $y^{\prime }=y/L$, $y_{0}^{\prime }=y_{0}/L$
$t^{\prime } = \gamma t$, $l^{\prime } = l/L$, $\omega_{D}^{\prime }=\omega _{D}/\gamma$,  
$F^{\prime }=F/mL\gamma^{2}$,  
$F_{D}^{\prime }=F_{D}/mL\gamma^{2}$, $V_{r}^{\prime } = V_{r} /mL^{2}\gamma^{2}$,
$V_{1}^{\prime } = V_{1} /mL^{2}\gamma^{2}$,and $V_{b}^{\prime } = V_{b} /mL^{2}\gamma^{2}$.
Thus, we are using the periodicity of the potential $L$ as the natural length scale and 
the inverse of the friction coefficient $\gamma$ defines the natural time scale. With these
two quantities, the natural force is given by $mL\gamma^{2}$ and the associated energy
by  $mL^{2}\gamma^{2}$.

The dimensionless equation of motion, after renaming the variables again without the 
primes, becomes 

\begin{equation}
\dot x = - {\frac{dV(x)}{dx}} - {\frac{\partial V_{b}(x-y)}{\partial x}} + \sqrt{2D}\xi_{1}(t) 
- F_D \cos(\omega_D t),
\end{equation}

\begin{equation}
\dot y = - {\frac{dV(y)}{dy}} - {\frac{\partial V_{b}(x-y)}{\partial y}} + \sqrt{2D}\xi_{2}(t)
+ F_D \cos(\omega_D t),
\end{equation}

\noindent where the dimensionless ratchet potential can be written as

\begin{equation}
V(x) = V_{1} - V_{r} \left [\sin (2\pi (x-x_0)) + {\frac{1}{4}}\sin (4\pi (x-x_0)) \right ],
\end{equation}

\noindent and the dimensionless bistable potential is

\begin{equation}
V_{b}(x-y) = V_{b} + V_{b} \left [{\frac{(x-y)^4}{l^4}} - 2{\frac{(x-y)^2}{l^2}} \right ].
\end{equation}

It is convenient to rewrite the equations of motion in terms of an effective time-dependent
bistable potential that incorporates the periodic forcing as

\begin{equation}\label{emov1}
\dot x = - {\frac{dV(x)}{dx}} - {\frac{\partial U_{b}(x-y,t)}{\partial x}} + \sqrt{2D}\xi_{1}(t),
\end{equation}

\begin{equation}\label{emov2}
\dot y = - {\frac{dV(y)}{dy}} - {\frac{\partial U_{b}(x-y,t)}{\partial y}} + \sqrt{2D}\xi_{2}(t),
\end{equation} 

where the potential $U_{b}$ is given by

\begin{equation}
U_{b}(x-y,t) =V_b(x-y)+ (x-y)F_D \cos(\omega_D t).
\end{equation}

In this way, it is clear that the dynamics corresponds to a rocking bistable potential
for the relative coordinate $x-y$. 
 
In Fig. 1, we show the walker on the ratchet potential. The ``feet'' of the
walker are indicated by the particles at the coordinates $x$ and $y$. In the inset we depict the 
bistable potential that couples both particles. This model, at variance with many others
that consider only a linear coupling, incorporates a nonlinear coupling between the
two particles, as has been discussed before \cite{mat2005}. The bistable potential 
depends on the variable $x - y$, that can be positive, negative or zero. When 
$x - y > 0$, the $x$ particle is leading, and when $x - y < 0$, the $y$ particle is the 
leading one. Thus, the transitions between the two wells in the bistable potential 
correspond to an exchange of the order between the particles. The minima are
located at $x - y = \pm 1$, and correspond 
to the two stable equilibrium configurations for the walker; on the other hand,
the maximum at the origin ($x - y = 0$) is unstable. So, we can think of a state 
oscillating in the bistable potential back and forth between the two minima, as the 
walker alternating theirs two feet.
In Fig. 1a, we show the case when $y$ (red or black) is larger that $x$ (green or gray)
and in Fig. 1b, we depict the opposite situation after a single step.
The walker is moving forward steeping on the ratchet potential alternating the ordering
of their feet. This systematic movement corresponds to a synchronized oscillation of the
internal degree of freedom $x-y$ on the rocking bistable potential. See the insets in
Fig. 1.     

\begin{figure}
\begin{center}
\includegraphics[width=8.0cm]{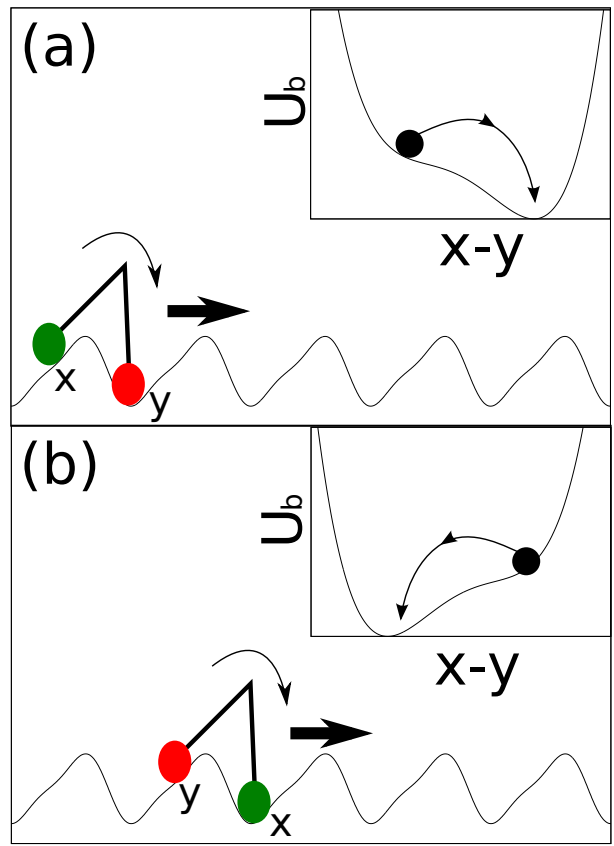}
\end{center}
\caption{The walker on the ratchet potential. 
In (a) we show the case when $y$ (red or black) is larger that $x$ (green or gray)
and in (b) we depict the opposite situation after a single step.
In the inset we show the rocking bistable potential 
as a function of the distance $x-y$; (a) for $x-y<$0, the foot $x$ (green or gray) is behind the 
foot $y$ (red or black) and, (b) for $x-y>$0, the foot $y$ is behind the foot $x$. When the walker 
makes a step by alternating the positions of the feet, that corresponds to a transition 
between the wells of the bistable potential.}
\label{fig1}
\end{figure}

\section{Numerical results}

In this section we will solve numerically, using the fourth-order Runge-Kutta 
algorithm, the dynamical system given by  Eqs. (\ref{emov1}, \ref{emov2}) for the walker on a ratchet.
We used the Fox algorithm \cite{fox}, that integrates the deterministic part of the
equations using a fourth-order Runge-Kutta algorithm, whereas the stochastic
part is integrated with a Taylor scheme; this method is called the exact propagator \cite{man}.

Once we solve the system, we calculate the position of each of the particles of the walker. 
We will fix throughout the paper the following parameters: the amplitude of the ratchet 
potential as $V_r = 1/2 \pi$, the amplitude of the bistable potential as $V_b = V_r$, that is, 
both amplitudes are equal; we fix $l = 0.5$, $F_D = 1$ and $\omega_D = 0.1$. 
The initial conditions are $x(t=0)=0$ and $y(t=0)=1$. 

The dynamics of the walker can be describe as two coupled particles on a one-dimensional ratchet 
potential, or as a single particle in an effective time-dependent two-dimensional potential 
$U(x,y,t)$, given by 
\begin{equation}
 U(x,y,t)=V(x)+V(y)+U_b(x-y,t). 
\end{equation}

In Fig. 2, we show a three-dimensional plot of this 
effective potential $U(x,y,t)$ as a function of $x$ and $y$, fixing the time as $t=T$, where $T=2\pi/\omega_D$, and the contour lines at the bottom. Due to the coupling between the particles, 
the effective potential resemble a channel with maxima and minima. Since this
potential varies in time, the entire profile changes accordingly in a periodic fashion. 
In Fig. 3, we show the 2D contour map and a typical trajectory when the noise intensity is $D = 0.02$. 
When this trajectory crosses the diagonal line $x=y$, the walker performs a step that
exchange the order of the two feet. It is clear from this trajectory that the walker is stepping
through the potential in a systematic and synchronized way. 
For clarity, in Fig. 4, we show the same trajectory
depicted in Fig. 3, using the same parameters, but this time showing the dynamics of
the two particles $x$ and $y$ as a function of time. Notice here that there is a systematic
positive current where the order of the feet exchange periodically on average.   

\begin{figure}
\begin{center}
\includegraphics[width=8.0cm]{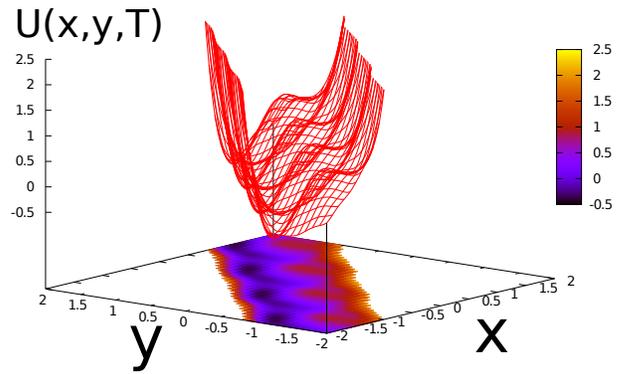}
\end{center}
\caption{The effective time-dependent 3D potential $U(x,y,t)=V(x)+V(y)+U_b(x-y,t)$ 
as a function of $x$ and $y$, fixing the time as $t=T$, where $T=2\pi/\omega_D$. 
At the bottom of the graph we show the contour map of this 3D graph.}
\label{fig2}
\end{figure}

\begin{figure}
\begin{center}
\includegraphics[width=8.0cm]{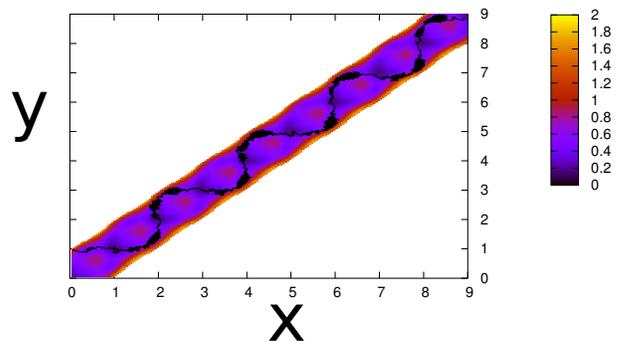}
\end{center}
\caption{We show the same contour map as in figure 2 and a typical trajectory of 
the particle over this effective potential, for the noise intensity $D=0.02$. 
When the trajectory crosses the diagonal line $x=y$, the walker performs a step
that exchange the order of the two feet.}
\label{fig3}
\end{figure}

\begin{figure}
\begin{center}
\includegraphics[width=8.0cm]{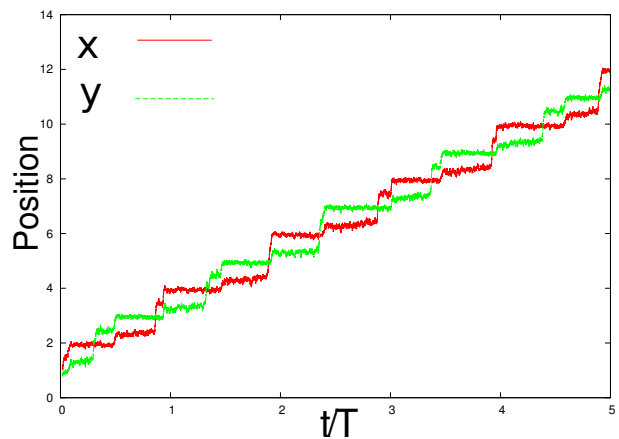}
\end{center}
\caption{We show the same trajectories as in figure 3, using the same parameters,
for the two feet of the walker as a function of time.}
\label{fig4}
\end{figure}

In Fig. 5, we show the internal degree of freedom $(x-y)/l$ as a function of time. We scale 
this internal degree freedom with the characteristic length $l$ of the bistable potential. The 
green (dashed) line corresponds to the deterministic case, without noise $(D = 0)$,
where the dynamics is confined and oscillates around the minimum at $x-y = -l$ with 
the period $T = 2\pi/\omega_D$. In this case, the velocity of the walker is zero
and there is no net transport despite the presence of the asymmetric ratchet potential (see Fig. 6).
However, in the presence of a small amount of noise with $D = 0.02$, we have a
completely different behavior, as depicted in the red (full) line. Now, assisted by the noise,
the internal degree of freedom fluctuates around each of the two minima of the bistable potential
and jumps periodically from one minimum to the other, surmounting the potential
barrier at $x-y=0$. Notice that the average periodicity of this stochastic dynamics is precisely
the same periodicity of the external driving force $T = 2\pi/\omega_D$. In this case,
thanks to the presence of noise, the average velocity of the walker
is different from zero and we obtain a net transport through the ratchet (see Fig. 6).Thus, the dynamics
depicted in this figure illustrates very clearly the phenomenon of stochastic resonance
for the internal dynamics and its connection with the ratchet effect. 

\begin{figure}
\begin{center}
\includegraphics[width=8.0cm]{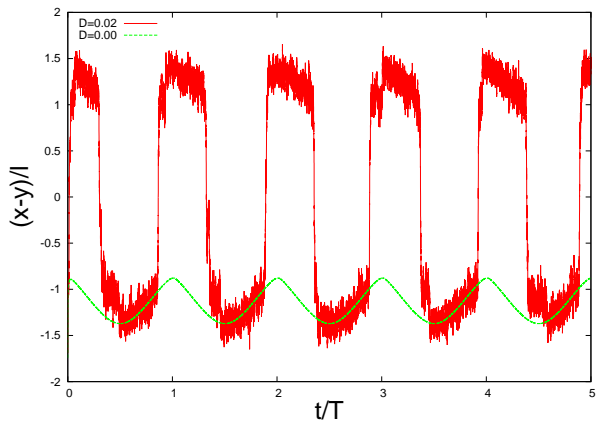}
\end{center}
\caption{The internal degree of freedom of the walker $(x-y)/l$ as a function of time.
The green (dashed) line shows the deterministic case without noise ($D=0$), 
and the red (full) line indicates the case when a noisy signal is acting on 
the walker with an intensity $D=0.02$. The signature of stochastic resonance is very
clear.}
\label{fig5}
\end{figure}

The central result of this paper is depicted in Fig. 6, where we show two quantities
that characterized two different phenomena: stochastic resonance and the ratchet effect.
The latter is characterized by the rectification, due to symmetry breaking, of external 
unbiased forces, giving as a result a net transport (current) or a finite average velocity. 
On the other hand, stochastic resonance is characterized by a maximum in the 
periodic response of the system to a periodic input signal, as a function of noise. 

Let us define these two quantities in more detail. The current is defined as the average
velocity of the center of mass of the walker. The center of mass is simply $(x+y)/2$ and
knowing $x(t)$ and $y(t)$ we calculate the asymptotic velocity of the center of mass. Then we 
perform an ensemble average over 1500 realizations of the noise. In this way we obtain
the average velocity of the center of mass $\langle v \rangle$ and rescale this quantity as 
$2\pi \langle v \rangle /\omega_D$\cite{ala2006,ala2007}. This is the main quantifier for the ratchet effect of the
walker with two Brownian motors moving on the ratchet potential. In Fig. 6, we plot this
current as a function of the noise intensity $D$ and notice that in the limit of zero noise
(deterministic case) the current is zero; on the other hand, for a large intensity of noise the
current tends to zero. However, for a certain optimal amount of noise the current is maximum.
See the red (black) line in Fig. 6.  In our case this maximum is located around $D = 0.02$.
The question is: Why at this particular noise? In order to answer this, we have to consider
the internal dynamics given by the internal degree of freedom of the walker. As we have
discussed before, the dynamics of this internal degree of freedom is given by a rocking bistable potential
in the presence of noise, that is, the equation of motion corresponds to an overdamped
particle in a double well potential tilted with a periodic function $F_D \cos(\omega_D t)$.
The mean value of the response is obtained by averaging over an ensemble of noise 
realizations (1500 in our case) and for small amplitudes this response is also periodic,
with the same period, and can be written as \cite{revsr3}

\begin{equation}
\langle (x-y)(t) \rangle =  \langle A \rangle \cos(\omega_D t - \langle \phi \rangle),
\end{equation}
with an amplitude $\langle A \rangle$ and a phase lag $\langle \phi \rangle$.
In Fig. 6, we show as a green (dashed) line the amplitude $\langle A \rangle /l$
as a function of the intensity of noise $D$. As expected, we obtain a maximum
that is a clear signature of stochastic resonance. Not only that, the optimal noise 
is precisely around $D = 0.02$. This means that, at this value, the internal degree of freedom
is jumping periodically between the two minima in the bistable potential, that correspond
to a synchronized alternation of the two feet of the walker. In summary, the conclusion
is that the synchronized walking due to stochastic resonance corresponds to an
optimal transport of the walker. In this way, we show the strong link between 
stochastic resonance and the ratchet effect. 

\begin{figure} 
\begin{center}
\includegraphics[width=8.0cm]{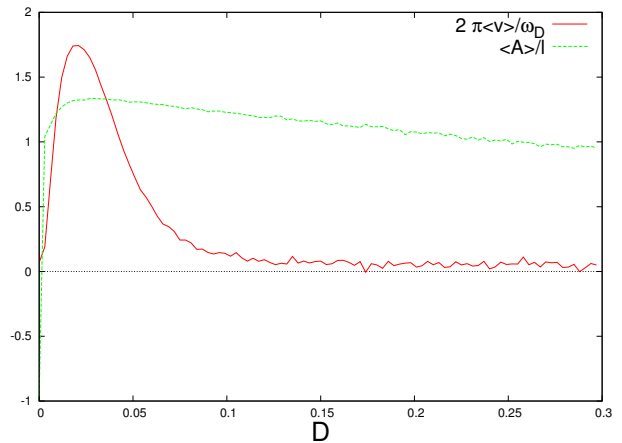}
\end{center}
\caption{The red (full) line shows the scaled average velocity $2\pi \langle v \rangle /\omega_D$
as a function of noise $D$. The green (dashed) line depicts the amplitude 
$\langle A \rangle /l$ as a function of noise $D$. Notice that the two maxima coincide at
the same amount of optimal noise around $D = 0.02$, indicating the strong connection
between stochastic resonance and optimal transport.}
\label{fig6}
\end{figure}

In Fig. 7, we show the same as in Fig. 6, but for a periodic symmetric potential
of the form $\cos(2\pi x)$, instead of an asymmetric ratchet potential. In this case we still 
have the effect of stochastic resonance for the internal degree of freedom, but now the current 
is zero (with some fluctuations due to the finiteness of the ensemble of realizations of noise). 
Thus, even though the walker is alternating their feet, it is not moving systematically in any direction since 
there is no symmetry breaking in the potential that rectifies the motion.
A similar case of a neutral dipole in a symmetric 1D substrate has been studied by \cite{march2010}.

\begin{figure}
\begin{center}
\includegraphics[width=8.0cm]{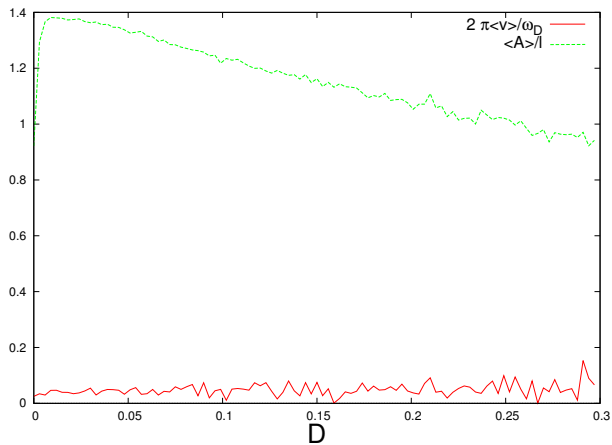}
\end{center}
\caption{The same two quantities as in Fig. 6, but instead of an asymmetric ratchet
potential, we used a symmetric potential of the form $\cos(2\pi x)$.}
\label{fig7}
\end{figure}

\section{Concluding remarks}

We have analyzed the stochastic dynamics of a walker comprising two particles (Brownian motors)
coupled through a bistable potential. This double-well potential is rocked by a periodic force
in the presence of thermal noise, and thus is able to manifest the phenomenon of stochastic
resonance. The walker is moving on a one-dimensional asymmetric ratchet potential by 
interchanging the order of the two particles and an ensemble of these walkers show an average directed
current in the presence of noise. In this paper we show first that the internal degree of freedom of
the walker exhibits stochastic resonance and then we show that the average current has a
maximum for an optimal amount of noise. On the other hand, stochastic resonance is 
characterized by the amplitude of the periodic response as a function of noise;
this amplitude shows a maximum for a particular value of the noise intensity. As we demonstrate
in this paper, the maximum of the current and the maximum of the stochastic resonance occurs
at the same amount of noise. Therefore, the main conclusion is that the synchronized walking 
due to stochastic resonance corresponds to an optimal transport of the walker. 
In this way, we show the intimate relation between stochastic resonance and the ratchet effect.

\bigskip
{\bf Acknowledgements}
\medskip

 FRA gratefully acknowledges financial support from C3-UNAM posdoctoral fellowship.

\end{document}